\documentstyle[12pt,epsf]{article}

\newcommand{\bea}{\begin{eqnarray}}
\newcommand{\eea}{\end{eqnarray}}
\newcommand{\beq}{\begin{equation}}
\newcommand{\eeq}{\end{equation}}
\newcommand{\nn}{\nonumber}

\def\k{{\vec k}}
\def\x{{\vec x}}
\def\/{\over}
\begin{document}

\parindent=1 em
\frenchspacing

\title{
Radiative energy shifts of accelerated atoms}
\author{
\large J\"urgen Audretsch\thanks{e-mail: Juergen.Audretsch@uni-konstanz.de}
 and Rainer M\"uller\thanks{e-mail: Rainer.Mueller@.physik.uni-muenchen.de
\hfill {\it to appear in Phys. Rev. A}\quad}\\
\normalsize \it Fakult\"at f\"ur Physik der Universit\"at Konstanz\\
\normalsize \it Postfach 5560 M 674, D-78434 Konstanz, Germany}
\date{}
\maketitle
\vspace{5mm}

\begin{abstract}
We consider the influence of acceleration on the radiative energy shifts
of atoms in Minkowski space. We study a two-level atom coupled to
a scalar quantum field. Using a Heisenberg picture approach, we are
able to separate the contributions of vacuum fluctuations and radiation
reaction to the Lamb shift of the two-level atom. The resulting energy
shifts for the special case of a uniformly accelerated atom are then
compared with those of an atom at rest.
\end{abstract}
\vspace{1cm}

\section{Introduction}

One of the most remarkable effects predicted by quantum field theory
is the Lamb shift, the shift of the energy levels of an atom which is
caused by the coupling to the quantum vacuum. It is known that this
level shift can be modified by external influences like a cavity
\cite{Meschede92}, for example. Its presence alters the mode structure of
the vacuum and leads to a Lamb shift which is different from its
free-space value.

In this paper, we study the effect of acceleration on radiative
energy shifts. It may not seem obvious at first sight why acceleration
should lead to a modification of the Lamb shift. To see this, one has to
combine results from different subfields of physics. First we note that
for a uniformly accelerated observer, the Minkowski vacuum appears as a thermal
heat bath of so-called ``Rindler particles''. This is usually interpreted as
a consequence of the
non-equivalence of the particle concept in inertial and accelerated frames
\cite{Fulling73,Unruh76,Birrell82}. The second ingredient we need is
the fact that the presence of photons leads to additional energy shifts
for atomic systems. This effect is called AC Stark shift or light shift
\cite{Cohen-Tannoudji92} and is connected with the virtual absorption and
emission of real photons. In particular, a thermal photon field causes
such an effect \cite{Barton72,Knight72,Farley81}. Consequently, taking these
results together, we can gain a heuristic insight why the Lamb shift of an
atom is modified if the atom is uniformly accelerating. A corresponding
effect is to be expected for other types of acceleration.

The first aim of this paper is the
calculation of radiative energy shifts. It can be carried out in an
elegant manner using the formalism of Dalibard, Dupont-Roc, and
Cohen-Tannoudji (DDC) \cite{Dalibard82,Dalibard84}. This approach has
the advantage that it allows also to separate the
contributions of vacuum fluctuations and radiation reaction to the
energy shifts. The independent treatment of these two effects
has a tradition in Heisenberg picture quantum electrodynamics
\cite{Ackerhalt73,Senitzky73,Milonni73,Ackerhalt74,Milonni75}
and beyond \cite{Barut90}.
In a previous paper \cite{Audretsch94}, we have studied the influences
of vacuum fluctuations and radiation reaction on the spontaneous transitions
of a uniformly accelerated atom moving through the Minkowski vacuum
It leads to a modified value of the Einstein A coefficient for spontaneous
emission. In addition we have shown how the lack of a balance between
vacuum fluctuations and radiation reaction causes a spontaneous excitation
from the ground state. This gives an interpretation of the physics
underlying the Unruh effect. For the radiative shift of atomic levels, the
separate discussion of vacuum fluctuations and radiation reaction may
also be interesting from a conceptual point of view, since in heuristic
pictures the Lamb shift has been often associated with the notion of
vacuum fluctuations alone \cite{Welton48}. This discussion is the second
aim of this paper.

The fully realistic calculation of the influence of acceleration on the Lamb
shift would require to deal with a multilevel atom coupled to the
electromagnetic field. This will not be done in the present paper.
To keep the discussion as clear and transparent as possible, we will
restrict to the simplest nontrivial example, a two-level atom in
interaction with a massless scalar quantum field. However, as we will see,
most of the essential features of the full problem are also present in
the simple model. The structure of the results can be seen clearly.
Furthermore, we will discuss only the nonrelativistic contribution
to the energy shift, i. e. we neglect the effects which are due to
the quantum nature of the electron field. However, it is well known
that the nonrelativistic part gives the dominant contribution to the
Lamb shift. Since we are only interested in the structure of the
results, we will concentrate on this part of the problem.

Our treatment applies the formalism of DDC to the case of a two-level
atom coupled to a scalar quantum field and generalizes it to an
arbitrary stationary trajectory of the atom. We will proceed as
follows: We consider the time evolution of an arbitrary atomic observable
$G$ as given by the Heisenberg equations of motion. Since we are only
interested in atom variables, we trace out the field degrees of freedom
in the part of the Heisenberg equations that is due to the coupling with
the field. It is then possible to identify in the resulting expressions
unambiguously the part that acts as an effective Hamiltonian with respect
to the time evolution of $G$. The expectation value of this operator
in an atom state $|b \rangle$ gives the radiative energy shift of this
level.

We must emphasize that the modification of radiative level shifts by
acceleration which is considered here must carefully be distinguished
from the direct influence of acceleration or curvature on the energy levels
of an atom \cite{Parker80,Audretsch93}. These corrections are obtained
in the simplest case by the inclusion of a term $amx$ into the Dirac
equation of the atom. They are assumed to be already included in the
otherwise unperturbed energies $\pm {1\/ 2} \omega_0$ of the atom.
In contrast to this, the effects considered in
this paper are true radiative corrections caused by the interaction of
the atom with the quantum field.

The organization of the paper is the following: In Sec. 2, we define
our model and set up the Heisenberg equations of motion, which are
solved formally. We apply the formalism of DDC to our model in Sec. 3
and generalize it to an atom moving on an arbitrary stationary trajectory
through the Minkowski vacuum.
In Sec. 4, we calculate the level shift for a two-level atom at rest.
We find that for the symmetric operator ordering adopted here, the only
contribution to relative energy shifts comes from vacuum fluctuations,
while the effect of radiation reaction is the same for both levels.
The result obtained for the scalar theory are then compared to the
standard results for the electromagnetic field and a similar structure
is found. Sec. 5 deals with a uniformly accelerated atom. Because the
analysis becomes more involved in this case, we use some methods from
quantum field theory in accelerated frames, which are discussed in the
Appendix. As a result, we find that the contribution of vacuum fluctuations
to the level shift is altered by the appearance of a thermal term with
the Unruh temperature $T= \hbar a/(2\pi c k)$. The contribution of
radiation reaction
is the same as for an atom at rest and does not contribute to relative
shifts. Finally, we point out the similarity of the results to those
obtained for the Lamb shift in a thermal heat bath
\cite{Barton72,Knight72,Farley81}.

\section{Two-level atom interacting with a massless scalar quantum field}

To investigate how the radiative energy shifts of atoms are modified by
acceleration, we choose the simplest nontrivial example: a two-level
atom in interaction with a real massless scalar field. We consider an
atom on a timelike Killing trajectory $x(\tau) =( t(\tau), \x(\tau))$, which
is parametrized by the proper time $\tau$. It will be called {\it stationary
trajectory} \cite{Letaw81}. The important consequence of the stationarity
is that the level spacing $\omega_0$ of the two states $|- \rangle$ and
$|+ \rangle$ of the atom does not depend on $\tau$. The zero of energy
is chosen so that that the energies of the two stationary states are
$-{1\/2}\omega_0$ and $+{1\/2}\omega_0$. As mentioned in the Introduction,
a possible constant modification of the energy level spacing which is
directly caused by the acceleration is assumed to be already included.
The free Hamiltonian of the two-level atom which generates the
atom's time evolution with respect to the proper time $\tau$ is given by
\beq H_A (\tau) =\omega_0 R_3 (\tau) \label{eq1} \eeq
where we have written $R_3 = {1\/2} |+ \rangle \langle + | - {1\/2}
|- \rangle \langle - |$, following Dicke \cite{Dicke54}.

The free Hamiltonian of the quantum field is
\beq {\tilde H}_F (t)  = \int d^3 k \, \omega_\k\, a^\dagger_\k a_\k .
\label{eq2}\eeq
where $a^\dagger_\k$, $a_\k$ are creation and annihilation operators
`photons' with momentum $\k$. The Hamiltonian (\ref{eq2}) governs the
time evolution of the quantum field with respect to the time variable $t$
of the inertial frame. However, to derive the Heisenberg equations of
motion of the coupled system, we have to choose a common time variable.
It turns out that it is most reasonable to refer generally to the atom's
proper time $\tau$. The free Hamiltonian of the field with respect to
$\tau$ can be obtained by a simple change of the time variable in the
Heisenberg equations:
\beq H_F (\tau) = \int d^3 k\, \omega_\k \,a^\dagger_\k a_\k
	{dt\/d \tau}. \label{eq3}\eeq

We couple the atom and the quantum field by the interaction Hamiltonian
\beq H_I = \mu R_2 (\tau) \phi ( x(\tau)) . \label{eq5}\eeq
which is a scalar model of the electric dipole interaction. Here we
have introduced $R_2 = {1\/2} i ( R_- - R_+)$, where $R_+ = |+ \rangle
\langle - |$ and $R_- = |- \rangle \langle +|$ are raising and lowering
operators for the atom. $\mu$ is a small coupling constant. The field
operator in (\ref{eq5}) is evaluated along the world line $x(\tau)$ of
the atom.

In the solutions of the Heisenberg equations of the atom and field
variables, two physically different
contributions can be distinguished: (1) the {\it free part}, which is
the part of the solution that goes back to the free Hamiltonians (\ref{eq1})
and (\ref{eq3}) and which is present even in the absence of the interaction,
(2) the {\it source part} which is caused by the coupling between atom
and field and represents their mutual influence:
\bea R_\pm (\tau) &=& R^f_\pm (\tau) + R^s_\pm (\tau),  \\
	R_3(\tau) &=& R^f_3 (\tau) + R_3^s (\tau),  \\
	\phi(x(\tau)) &=& \phi^f (x(\tau)) + \phi^s (x(\tau)).\label{eq5a}\eea
Their explicit form has been given in \cite{Audretsch94}.

\section{Radiative energy shifts: The contributions of vacuum fluctuations
and radiation reaction}

To determine the Lamb shift of an accelerated atom, we use
the physically appealing formalism of Dalibard, Dupont-Roc,
and Cohen-Tannoudji (DDC) \cite{Dalibard82,Dalibard84}. We will generalize
it for calculating level shifts to an atom on an arbitrary
stationary world line $x(\tau)$. One attractive feature of the
formalism is that it allows a separate
discussion of the two physical mechanisms which both contribute to
radiative energy shifts: the contributions of {\it vacuum fluctuations} and
of {\it radiation reaction}. Let us first discuss how these two parts
can be separated (for a more detailed discussion see \cite{Audretsch94}):
Because of the coupling (\ref{eq5}), the field operator
$\phi$ will appear in the Heisenberg equation of motion of an
arbitrary atomic variable $G$. According to (\ref{eq5a}), $\phi$
can be split into its free and source part. The following physical
mechanisms can be connected with these two contributions:
(1) the part of the Heisenberg equation which contains $\phi^f$
represents the change
in G due to {\it vacuum fluctuations}. It is caused by the fluctuations
of the field which are present even in the vacuum. (2) The term
containing $\phi^s$ represents the influence of the atom on the field.
This part of the field in turn reacts back on the atom.
This mechanism is called {\it radiation reaction}. Explicitly, the
contributions of vacuum fluctuations and radiation reaction to
$dG/d \tau$ are:
\beq \left( {d G(\tau) \/ d \tau} \right)_{vf} = {1\/2}i \mu \Bigl( \phi^f
	(x(\tau)) [R_2 (\tau), G (\tau)] + [R_2 (\tau), G (\tau)]
	\phi^f (x(\tau)) \Bigr), \label{eq15}\eeq
\beq \left( {d G(\tau) \/ d \tau} \right)_{rr} = {1\/2}i \mu \Bigl( \phi^s
	(x(\tau)) [R_2 (\tau), G (\tau)] + [R_2 (\tau), G (\tau)]
	\phi^s (x(\tau)) \Bigr), \label{eq16}\eeq
where we followed DDC \cite{Dalibard82} in choosing a symmetric ordering
between atom and field operators. Because we are interested only in
observables of the atom, we take the average of (\ref{eq15})
and (\ref{eq16}) in the vacuum state of the quantum field. In a
perturbative treatment, we take into account only terms up to
second order in $\mu$. Proceeding in a similar manner as in Refs.
\cite{Dalibard84,Audretsch94}, it is then possible to identify in
Eqs. (\ref{eq15}) and (\ref{eq16}) an {\it effective Hamiltonian}
for atomic observables that governs the time evolution with respect to
$\tau$ in addition to the free hamiltonian (\ref{eq1}). The averaged
equations (\ref{eq15}) and (\ref{eq16}) can be written
\beq \langle 0| \left( {d G (\tau) \/ d\tau} \right)_{vf,rr} |0\rangle =
	i [H^{eff}_{vf,rr} (\tau), G(\tau)] + \hbox{non-Hamiltonian terms}.
	\label{eq26}\eeq
where in order $\mu^2$
\beq H_{vf}^{eff} (\tau) = {1\/ 2} i \mu^2 \int_{\tau_0}^\tau d \tau'
  C^F(x(\tau),x(\tau')) [R_2^f(\tau'),R_2^f(\tau)] \label{eq33}\eeq
\beq H_{rr}^{eff} (\tau) = -{1\/ 2} i \mu^2 \int_{\tau_0}^\tau d \tau'
  \chi^F(x(\tau),x(\tau')) \{ R_2^f(\tau'),R_2^f(\tau)\} \label{eq34}\eeq
These effective Hamiltonians depend on the field variables only through
the simple statistical functions
\beq C^{F}(x(\tau),x(\tau')) = {1\/2} \langle 0| \{ \phi^f (x(\tau)), \phi^f
	(x(\tau')) \} | 0 \rangle, \label{eq24} \eeq
\beq \chi^F(x(\tau),x(\tau')) = {1\/2} \langle 0| [ \phi^f (x(\tau)), \phi^f
	(x(\tau'))] | 0 \rangle \label{eq25}\eeq
which are called the {\it symmetric correlation function} and the {\it linear
susceptibility of the field}.
The non-Hamiltonian terms in (\ref{eq26}) describe the effects of relaxation.

The expectation values of the effective Hamiltonians (\ref{eq33}) and
(\ref{eq34})
in an atomic state $|b \rangle$ represent the energy shift of the atomic
level $|b\rangle$ caused by the coupling to the quantum field. The total
shift contains the contribution of vacuum fluctuations as well as the
contribution of radiation reaction. Taking the expectation values of
(\ref{eq33}) and (\ref{eq34}), we obtain the radiative energy shifts of the
 level
$|b\rangle$ due to vacuum fluctuations
\beq (\delta E_b)_{vf} = - i \mu^2 \int_{\tau_0}^\tau d \tau' \,
	C^F(x(\tau),x(\tau')) \chi^A_b(\tau,\tau'), \label{eq35} \eeq
and due to radiation reaction
\beq (\delta E_b)_{rr} = - i \mu^2
	\int_{\tau_0}^\tau d \tau' \, \chi^F(x(\tau),x(\tau'))
	C^A_b(\tau,\tau'), \label{eq36}\eeq
where we have defined the {\it symmetric correlation function}
\beq C^{A}_b(\tau,\tau') = {1\/2} \langle b| \{ R_2^f (\tau), R_2^f (\tau')\}
	| b \rangle \label{eq38}\eeq
and the {\it linear susceptibility of the atom}
\beq \chi^A_b(\tau,\tau') = {1\/2} \langle b| [ R_2^f (\tau), R_2^f (\tau')]
	| b \rangle \label{eq37}\eeq
As a result, we note that
Eqs. (\ref{eq35}) and (\ref{eq36}) for the radiative energy shifts
for atoms that move on an arbitrary stationary trajectory differs from
the results for atoms at rest \cite{Dalibard82,Dalibard84} only in that
the statistical functions of the field (\ref{eq24}) and (\ref{eq25}) are
evaluated along the world line of the atom which may now be an accelerated one.
The statistical functions of the atom do not depend on the trajectory
$x(\tau)$.

Below we will need the explicit forms of the different statistical functions.
Those refering to the atom can be written
\bea C^{A}_b(\tau,\tau') & =& {1\/2} \sum_d |\langle b| R_2^f (0) | d
	\rangle |^2 \left( e^{i \omega_{bd}(\tau - \tau')} + e^{-i \omega_{bd}
	(\tau - \tau')} \right), \label{eq39}\\
\chi^A(\tau,\tau')_b & =& {1\/2} \sum_d |\langle b| R_2^f (0) | d \rangle |^2
	\left(e^{i \omega_{bd}(\tau - \tau')} - e^{-i \omega_{bd}(\tau -
\tau')}
	\right), \label{eq40}\eea
where $\omega_{bd}= \omega_b-\omega_d$ and the sum extends over a complete
set of atomic states.

\section{Radiative energy shift for an atom at rest}

Let us first reproduce the standard result for the Lamb shift for
an atom at rest in the scalar theory. It can be compared afterwards
with the corresponding
result for a uniformly accelerated atom. The statistical functions of
the field for the trajectory
\beq t(\tau) =\tau, \qquad \x (\tau)=0 \label{eq41}\eeq
can be easily evaluated. We obtain
\bea
C^F(x(\tau),x(\tau')) &=& {1\/ 8\pi^2}\int d \omega \, \omega
	\left( e^{-i \omega
    (\tau-\tau')} + e^{i \omega (\tau-\tau')}\right) \label{eq42} \\
\chi^F(x(\tau),x(\tau')) &=& {1\/ 8\pi^2}\int d \omega\, \omega
	\left( e^{-i \omega
  	(\tau-\tau')} - e^{i \omega (\tau-\tau')}\right). \label{eq43}
\eea

The contribution of vacuum fluctuations to the radiative shift of level
 $|b\rangle$ can be calculated from (\ref{eq35})
\beq (\delta E_b)_{vf} = {\mu^2\/ 8\pi^2} \sum_d |\langle b|R_2^f(0)|
     d\rangle|^2 \int_0^\infty d\omega \,\omega \left({{\cal P}\/ \omega +
     \omega_{bd}} - {{\cal P}\/ \omega -\omega_{bd}} \right),
     \label{eq47}\eeq
where $\cal P$ denotes the principle value.
The integral is logarithmically divergent, as expected for a nonrelativistic
calculation of radiative shifts \cite{Bethe47}. As is well known, the
introduction of a cutoff
is therefore necessary. The summation over $d$ displays the role of virtual
transitions to other levels.

The {\it relative} energy shifts of the two levels due to vacuum fluctuations
can be calculated by evaluating the $d$ summation for $|b\rangle =
|\pm\rangle$. The resulting expression is
\bea \Delta_{vf} &=& (\delta E_+)_{vf} - (\delta E_-)_{vf} \\
     &=& {\mu^2\/ 16\pi^2} \int_0^\infty d\omega \,\omega \left({{\cal P}\/
 \omega
     +\omega_{bd}} - {{\cal P}\/ \omega -\omega_{bd}} \right).
     \label{eq48}\eea

For the contribution of radiation reaction we obtain
\beq (\delta E_b)_{rr} = -{\mu^2\/ 8\pi^2} \sum_d |\langle b|R_2^f(0)|
     d\rangle|^2 \int_0^\infty d\omega \,\omega \left({{\cal P}\/ \omega +
     \omega_{bd}} + {{\cal P}\/ \omega -\omega_{bd}} \right).
     \label{eq49}\eeq
This expression diverges linearly. However, as can be seen by explicitly
evaluating the sum, radiation reaction does not give a contribution to
the relative shift of the two levels:
\beq \Delta_{rr} = (\delta E_+)_{rr} - (\delta E_-)_{rr} = 0.
     \label{eq50}\eeq
Thus the {\it Lamb shift} as the the relative radiative energy shift of the
two-level atom at rest is caused entirely by vacuum fluctuations:
\bea \Delta_0  &\equiv& \Delta_{vf} + \Delta_{rr} \nonumber\\
     &=& {\mu^2\/ 16\pi^2} \int_o^\infty d\omega \,\omega \left({{\cal P}\/
	 \omega +\omega_{bd}} - {{\cal P}\/ \omega -\omega_{bd}} \right).
     \label{eq51}\eea
This expression agrees structurally with the standard result for the
Lamb shift of a two-level atom
\cite{Ackerhalt73,Senitzky73,Milonni73,Ackerhalt74,Milonni75}. The
modifications are caused by the differences between the electromagnetic
and the scalar theory.

We mention that Welton's picture of radiative shifts \cite{Welton48}, who
tried to interpret the Lamb shift only in terms of vacuum fluctuations
conforms with the fact that the energy shift (\ref{eq51}) is caused
solely by vacuum fluctuations. This feature is a peculiarity of the simple
model of a two-level atom, however. For a real multilevel atom,
the contribution of radiation reaction will be different for different
levels, and a mass renormalization will become necessary.

\section{Radiative energy shifts for a uniformly accelerated atom}

Let us now consider the case of a uniformly accelerated two-level atom.
It moves on the trajectory
\beq t(\tau)={1\/ a}\sinh a \tau, \qquad z(\tau)={1\/ a}\cosh a \tau,
     \label{eq52}\eeq
$$ x(\tau)=y(\tau)=0, $$
where $a$ is the proper acceleration. The calculation of the statistical
functions of the field turns out to be much more complicated for an
accelerated atom than for an atom at rest. It is most convenient to
employ methods from the quantum field theory in accelerated frames.
In order to keep the discussion transparent, we have put the
calculation into the Appendix and simply quote the results here:
\bea C^F (x(\tau),x(\tau')) &=& {1\/ 8\pi^2} \int_0^\infty d\omega'
   \,\omega' \coth\left({\pi\omega'\/a}\right)\left( e^{-i \omega'
   (\tau-\tau')} + e^{i \omega'(\tau-\tau')}\right) \label{eq53}\\
\chi^F (x(\tau),x(\tau')) &=& {1\/ 8\pi^2} \int_0^\infty d\omega'
   \,\omega' \left( e^{-i \omega' (\tau-\tau')} -
   e^{i \omega'(\tau-\tau')}\right) \label{eq54}\eea
We see that the expression (\ref{eq54}) for the linear susceptibility of the
field is formally identical for an accelerated atom and an atom at rest.
However, with regard to the cutoff prescription to be given below, it is
important to note that $\omega'$ in (\ref{eq53}) and
(\ref{eq54}) denotes the energy in the accelerated frame, i. e. as measured
by an observer on the trajectory (\ref{eq52}).

The remaining calculation is straightforward. By substituting the
statistical functions (\ref{eq39}), (\ref{eq40}), (\ref{eq53}), and
(\ref{eq54})
into the general formulas (\ref{eq35}) and (\ref{eq36}) for the level
shifts, we find
\bea (\delta E_b)_{vf} &=& {\mu^2\/8\pi^2} \sum_d|\langle b|R^f_2 (0)|d
    \rangle|^2 \nonumber\\
&&\qquad\times \int_0^\infty d\omega'\,\omega' \coth\left({\pi\omega'\/a}
    \right) \hbox{Im}\left\{ \int_0^\infty du \left( e^{i(\omega'
    +\omega_{bd})u} - e^{i(\omega'-\omega_{bd})u}\right)\right\},
    \label{eq55}\\
(\delta E_b)_{rr} &=& -{\mu^2\/8\pi^2} \sum_d|\langle b|R^f_2 (0)|d
    \rangle|^2 \nonumber\\
&&\qquad\times\int_0^\infty d\omega'\,\omega'
    \hbox{Im}\left\{ \int_0^\infty du \left( e^{i(\omega'
    +\omega_{bd})u} + e^{i(\omega'-\omega_{bd})u}\right)\right\}.
    \label{eq56}\eea

The result for the radiative energy shift for a uniformly accelerated atom
can now be obtained by evaluating the integrals. The contribution of vacuum
fluctuations is
\beq (\delta E_b)_{vf} =  {\mu^2\/8\pi^2} \sum_d|\langle b|R^f_2 (0)|d
    \rangle|^2 \int_0^\infty d\omega'\,\omega'\left( {1} + {2\/
    e^{2\pi\omega'/a}-1}\right)\left({{\cal P}\/\omega'+\omega_{bd}}
    - {{\cal P}\/\omega'-\omega_{bd}}\right).  \label{eq57}\eeq
Comparing this formula to Eq. (\ref{eq47}) for an atom at rest, we note
that the acceleration-caused correction is additive and contains a
 characteristic thermal term
with the Unruh temperature $T = \hbar a/(2\pi kc)$.
On the other hand, the contribution of radiation reaction,
\beq (\delta E_b)_{rr} = -{\mu^2\/8\pi^2} \sum_d|\langle b|R^f_2 (0)|d
    \rangle|^2 \int_0^\infty d\omega'\,\omega' \left({{\cal
	P}\/\omega'+\omega_{bd}}
    + {{\cal P}\/\omega'-\omega_{bd}}\right)  \label{eq58}\eeq
is  exactly the same as for an atom at rest. The uniform acceleration
has no effect on the shift caused by radiation reaction and we find
again no relative energy shift: $\Delta_{rr}=0$.

Thus the {\it Lamb shift} $\Delta$ can be obtained from the contribution of
vacuum fluctuations (\ref{eq57}) by evaluating the summation over $d$ for
each of the two levels separately:
\bea \Delta &=& \Delta_{vf}= (\delta E_+)_{vf} - (\delta E_-)_{vf}\nn\\
	&=& {\mu^2\/ 16\pi^2}\int_0^\infty d \omega'\, \omega' \left[ {1} +
{2\/
    e^{2\pi\omega'/a}-1}\right]\left({{\cal P}\/\omega'+\omega_0}
    - {{\cal P}\/\omega'-\omega_0}\right). \label{eq59}\eea
{}From the two terms in square brackets, we can distinguish two contributions
in (\ref{eq59}). The first one has
the same functional form as $\Delta_0$ of (\ref{eq51}) for an atom at rest.
Because it is logarithmically divergent, the introduction
of a cutoff is necessary. The only sensible way to do this is to
impose the cutoff frequency in the rest frame of the atom. Since
$\omega'$ in (\ref{eq59}) is the frequency in the accelerated system,
this can be done quite naturally in the present formalism.
The second contribution in (\ref{eq59}) represents the modification of the
Lamb shift caused by the acceleration and contains the thermal term.
Inspection of the integral shows that this correction is finite. Equation
(\ref{eq59}) can thus be written
\beq \Delta = \Delta_0 + {\omega_0\mu^2 \/ 16\pi^2} \,G \left({a\/ 2\pi
	\omega_0} \right), \label{eq60}\eeq
where $\Delta_0$ is the Lamb shift (\ref{eq51}) for $a=0$ and $G(u)$ is
defined by
\beq G(u) = u \int_0^\infty dx {x\/ e^x -1} \left({{\cal P}\/ x+u^{-1}} -
	{{\cal P}\/ x-u^{-1}} \right). \label{eq61}\eeq
The evaluation of the integral must be done numerically. The result
is shown in Fig. 1. In the limit of small $u=a/(2\pi \omega_0)$,
(\ref{eq60}) can be approximated by
\beq \Delta = \Delta_0 + {a^2\mu^2 \/ 192 \pi^2}{1 \/ \omega_0}.
	\label{eq62}\eeq
The dependence on $a$ can be read off from Fig. 1. To estimate the order of
magnitude of the acceleration needed for an appreciable effect, we demand
$(\Delta -\Delta_0)/\omega_0 = \mu^2/(16 \pi)$ and therefore $G(u)=1$.
This amounts to $a \approx c \omega_0$, which gives $a \approx 10^{24}
\hbox{m/s}^2$ for an 1 eV transition.

\begin{figure}[t]
\epsfysize=8cm
\hspace{1.5cm}
\epsffile{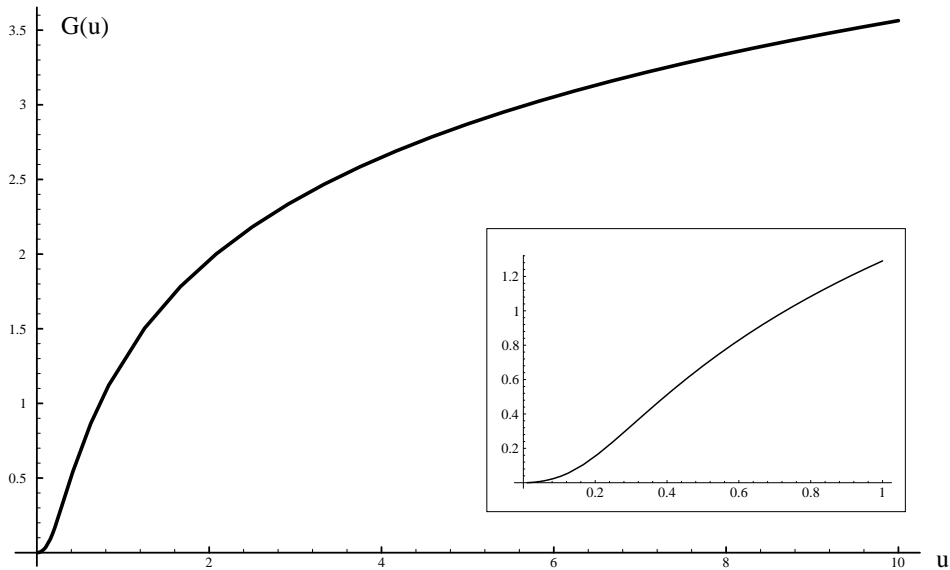}
\caption{
Plot of the function $G(u)$ defined in (\protect\ref{eq61}).
The inset shows the
behaviour at small values of $u$ (cf. (\protect\ref{eq62})).}
\end{figure}

Finally, we note the structural similarity of the equations (\ref{eq59})
and (\ref{eq60}) to the corresponding expressions obtained for the
Lamb shift in a thermal heat bath \cite{Barton72,Knight72,Farley81}.
They have in common the appearance of the factor in square brackets
under the integral in (\ref{eq59}) with $T= a/(2\pi)$. A
heuristic connection between the two physical situations has already been
given in the Introduction.
The differences are due to the properties of the scalar theory and the
fact that we considered a two-level atom instead of a real atom.

\section{Conclusion}

In this paper, we have discussed two main points. First, we studied the
influence of acceleration on the radiative energy shift of atoms. The
formalism can be applied to an atom on an arbitrary stationary trajectory.
We considered here the special case of uniformly accelerated motion.
The resulting expression for the Lamb shift of a uniformly accelerated
two-level atom is given in Eq. (\ref{eq59}) and (\ref{eq60}).
Comparison with the corresponding formula (\ref{eq51}) for an atom at rest
shows that the correction consists in the appearance of the factor in
square brackets, which contains the thermal term. This modification
is structurally the same as for the Lamb shift in a thermal heat bath.

The second main goal of the paper has been the identification of the two
physical mechanisms responsible for the radiative shifts. Using the
formalism of DDC, we were able to discuss the contributions of vacuum
fluctuations and radiation reaction separately. The effect of radiation
reaction is the same for a uniformly accelerated atom as for an atom at
rest. It does not affect relative energy shifts. It is interesting to note
that also in the case of other radiative phenomena like the Unruh effect and
spontaneous emission, radiation reaction is not influenced by acceleration
\cite{Audretsch94}. The contribution of vacuum fluctuation, however, is
modified by the thermal terms and gives the total effect.

\appendix

\section*{Appendix: Statistical functions of the field for a uniformly
accelerated atom}

The statistical functions of the field for an accelerated atom can be
most easily evaluated by using the well-known formalism of quantum
field theory in accelerated frames \cite{Fulling73}. For the details of
the theory, we refer to the literature \cite{Birrell82,Takagi86,Ginzburg87}.

The two Rindler wedges $R^+$ ($z > |t|$) and $R^-$ ($-z > |t|$) can
be covered by Rindler coordinates ($\eta, x', y', \xi$):
\beq t = {\epsilon\/a} e^{a\xi}\sinh a \eta, \qquad
     z = {\epsilon\/a} e^{a\xi}\cosh a \eta, \label{eqA1}\eeq
$$ x'=x, \qquad y'=y,$$
where $\epsilon=\pm$ refers to the wedge $R^\pm$. For $\xi=x'=y'=0$,
(\ref{eqA1}) reduces to (\ref{eq52}), describing a uniformly accelerated
observer
with acceleration $a$ on a trajectory with proper time $\eta$.

The metric of Minkowski spacetime takes in terms of the coordinates
(\ref{eqA1}) the following form
\beq ds^2 = e^{2a\xi} ( d\eta^2 - d\xi^2) - dx'^2 -dy'^2 \label{eqA2}\eeq
and the wave equation becomes
\beq \left[ e^{-2a\xi}(\partial^2_\eta - \partial^2_\xi) -
    \partial_{x'}^2 - \partial^2_{y'} \right] \phi(x) =0. \label{eqA3}\eeq
It can be shown that the normalized mode solutions which have positive
frequency with respect to the time variable $\eta$ are
\beq p^\epsilon_{\omega' k'_\perp} = {1\/ 2\pi^2\sqrt{a}} \sinh^{1\/ 2}
  ({\pi\omega'\/ a})\, K_{i{\omega'\/ a}}({1\/a} k'_\perp e^{a\xi})\,
  \exp( i \epsilon {\vec k'_\perp}{\vec x'_\perp} - i\epsilon \omega'
  \eta)  \label{eqA4}\eeq
in the wedge $R^\epsilon$. They are identically zero in the opposite
wedge. In (\ref{eqA4}), $\omega'$ denotes the energy with respect to
the time variable $\eta$, ${\vec x'_\perp}=(x',y')$, and
${k'}^2_\perp = {k'_x}^2 + {k'_y}^2$.
Quantization with respect to the mode functions (\ref{eqA4}) leads to the
concept of Rindler particles \cite{Fulling73}. However, Unruh
\cite{Unruh76} observed that the linear combinations
\beq v^\epsilon_{\omega' k'_\perp} = (1- e^{-2\pi\omega' /a})^{-{1\/2}}
   \left( p^\epsilon_{\omega' k'_\perp} + e^{-\pi\omega'/a}
   p^{\epsilon\ast}_{\omega' k'_\perp} \right) \label{eqA5}\eeq
have positive frequency with respect to the inertial time $t$.
Since they form a complete set in the wedges $R^\pm$, it is possible
to quantize the field using the Unruh modes (\ref{eqA5}).

The expansion of the field operators reads
\beq \phi (x) = \sum_\epsilon \int d\omega' \int d^2 k'_\perp
   \left( v^\epsilon_{\omega' k'_\perp}(x) b^\epsilon_{\omega' k'_\perp}
   + v^{\epsilon\ast}_{\omega' k'_\perp}(x)
   b^{\epsilon\dagger}_{\omega' k'_\perp} \right)\label{eqA6}\eeq
with creation and annihilation operators
$b^{\epsilon}_{\omega' k'_\perp}$, $b^{\epsilon\dagger}_{\omega' k'_\perp}$
which obey the usual commutation relations. The operators
$b^{\epsilon}_{\omega' k'_\perp}$ annihilate the Minkowski vacuum:
$b^{\epsilon}_{\omega' k'_\perp} |0 \rangle =0$.

It is now possible to calculate the statistical functions of the field
$C^F(x(\tau),x(\tau'))$ and $\chi^F(x(\tau),x(\tau'))$ as defined by
(\ref{eq24}) and (\ref{eq25}). To do this, we use the field expansion
(\ref{eqA6}). We stress, however, that no special physical status is
admitted to the Unruh modes (\ref{eqA5}). They serve only as a calculational
tool. Keeping this in mind, we consider the function
\beq \langle 0 | \phi(x(\tau)) \phi(x(\tau'))|0\rangle =
  \sum_\epsilon \int d\omega' \int d^2 k'_\perp
  v^{\epsilon}_{\omega' k'_\perp}(x(\tau))
  v^{\epsilon\ast}_{\omega' k'_\perp}(x(\tau')) \label{eqA7}\eeq
and evaluate the expression for the trajectory $\xi=x'=y'=0$,
$\tau=\eta'$. We find
\bea \langle 0 | \phi(x(\tau)) \phi(x(\tau'))|0\rangle &=&
  {1\/ 8\pi^4 a} \int d\omega' \int d^2 k'_\perp \, K^2_{i{\omega'\/a}}
\nonumber\\
&&\qquad\times
  ({1\/a}k'_\perp)\, \left( e^{\pi\omega'/a}e^{-i\omega'(\tau-\tau')}
  +e^{-\pi\omega'/a}e^{i\omega'(\tau-\tau')}\right). \label{eqA8}\eea
The statistical functions (\ref{eq24}) and (\ref{eq25}) can now be written
\bea C^F(x(\tau),x(\tau')) &=& {1\/8\pi^4 a} \int d^2 k'_\perp \int d\omega' \,
  K^2_{i{\omega'\/a}}
\nonumber\\
&&\qquad\times ({1\/a}k'_\perp) \cosh \left(
  {\pi\omega' \/a}\right) \left( e^{-i\omega'(\tau-\tau')}
  +e^{i\omega'(\tau-\tau')}\right), \label{eqA9}\\
 \chi^F(x(\tau),x(\tau')) &=& {1\/8\pi^4 a} \int d^2 k'_\perp \int d\omega' \,
  K^2_{i{\omega'\/a}}\nonumber \\
&&\qquad\times ({1\/a}k'_\perp) \sinh \left(
  {\pi\omega' \/a} \right)\left( e^{-i\omega'(\tau-\tau')}
  -e^{i\omega'(\tau-\tau')}\right). \label{eqA10}\eea
 After the evaluation of the integral
\beq \int d^2 k'_\perp \,  K^2_{i{\omega'\/a}} ({1\/a}k'_\perp)
   = {a\pi^2 \omega' \/ \sinh({\pi\omega'\/a})}, \label{eqA11}\eeq
where  we used Eq. (6.521.3) of Ref. \cite{Gradshteyn}, we obtain
the expressions (\ref{eq53}) and (\ref{eq54}) for the statistical functions
of the field for the uniformly accelerated atom.
\vskip 0.5cm

\noindent
{\bf Acknowledgments}\\
This work was supported by the Studienstiftung des deutschen Volkes.


\end{document}